\documentclass[a4paper,11pt]{article}

\usepackage{graphicx}
\usepackage{booktabs} 
\usepackage{subfigure} 
\usepackage{amssymb} 
\usepackage{wrapfig} 
\usepackage{xspace}
\usepackage[latin2]{inputenc}
\usepackage{indentfirst}
\usepackage{enumerate}
\usepackage{times}

\begin{document}
 
\begin{center}
{\bf \Large Hadron production measurement from NA61/SHINE}
\end{center}

\begin{center}
{A.Korzenev\footnote{Presented at NUFACT\,2013 in Beijing on Aug.23, 2013.}, on behalf of the NA61/SHINE collaboration}

{DPNC, University of Geneva, Switzerland}

{korzenev@mail.cern.ch}
\end{center}

\vspace*{0.3cm}

\begin{abstract}
New results from the NA61/SHINE experiment on the determination of
charged hadron yields in proton-carbon interactions are presented.
They aim to improve predictions of the neutrino flux
in the T2K experiment. 
The analysis is based on the main dataset collected by NA61/SHINE in the year 2009.
The data were recorded using
a secondary-proton beam of 31 GeV/$c$ momentum from CERN SPS
which impinges on a graphite target. 
To determine the inclusive production cross section for charged 
pions, kaons and protons a thin ($0.04\, \lambda_I$) target was exploited. 
Results of this measurement are used in the T2K beam simulation program
to reweight hadron yields at the interaction vertex.
At the same time,
NA61/SHINE results obtained with the T2K replica target ($1.9\, \lambda_I$)
allow to constrain hadron yields at the surface of the target. 
This allows to constrain up to 90\% of the neutrino flux,
thus reducing significantly the model dependence of the neutrino beam prediction.
All measured spectra are compared to predictions of hadron production models. 

\end{abstract}

\vspace{0.5cm}

In an accelerator neutrino experiment a precise knowledge of
a hadron production at the primary target is required \cite{review}. 
It is needed to predict the neutrino flux which is, in turn, used to calculate 
a neutrino cross section at the near detector, or provides an estimate of
the expected signal at the far detector for the study of neutrino oscillations.
In particular, the T2K experiment \cite{T2K_flux_paper} relies primarily on 
the hadron measurements performed by NA61 \cite{proposal_NA61} at CERN.
These data were taken by NA61 with a thin (0.04 $\lambda_{I}$) and 
with a full-size (1.9 $\lambda_{I}$) T2K replica targets 
in years 2007, 2009 and 2010.

For the cross section measurement NA61 used a secondary proton beam of 
31 GeV/$c$ momentum from CERN\,SPS scattered off a thin graphite target.
In the first physics analysis the pilot data collected in 2007 have been used.
The results on cross section of $\pi^\pm$ and K$^+$ \cite{Abgrall:2011ae,Abgrall:2011ts}
so far have been integrated to the T2K beam simulation program 
to constrain the production of hadrons in the primary interaction of 
the beam protons in the target \cite{T2K_flux_paper}.

\begin{figure}
\centering
\includegraphics[width=0.26\textwidth]{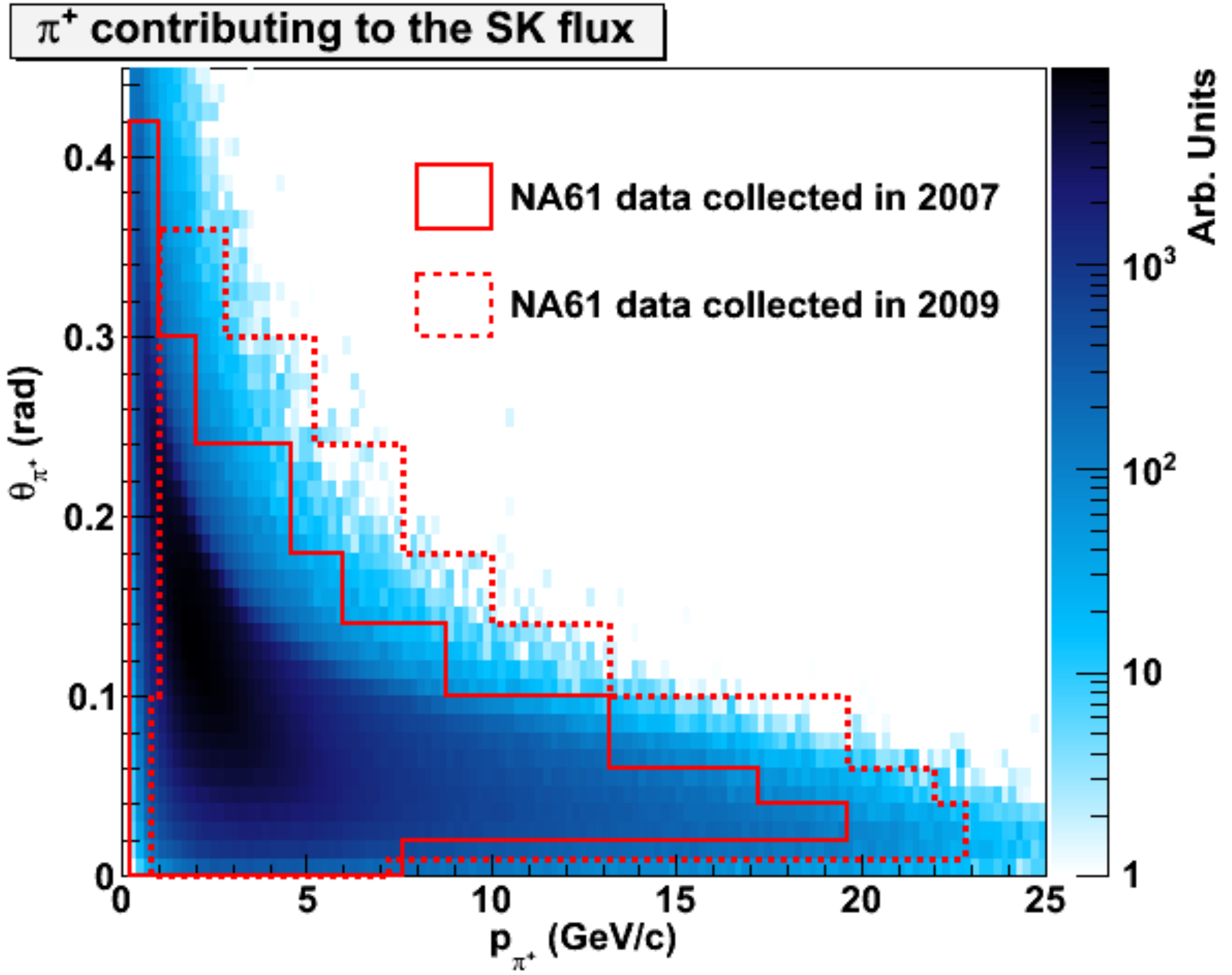}
\hspace*{0.4cm}
\includegraphics[width=0.26\textwidth]{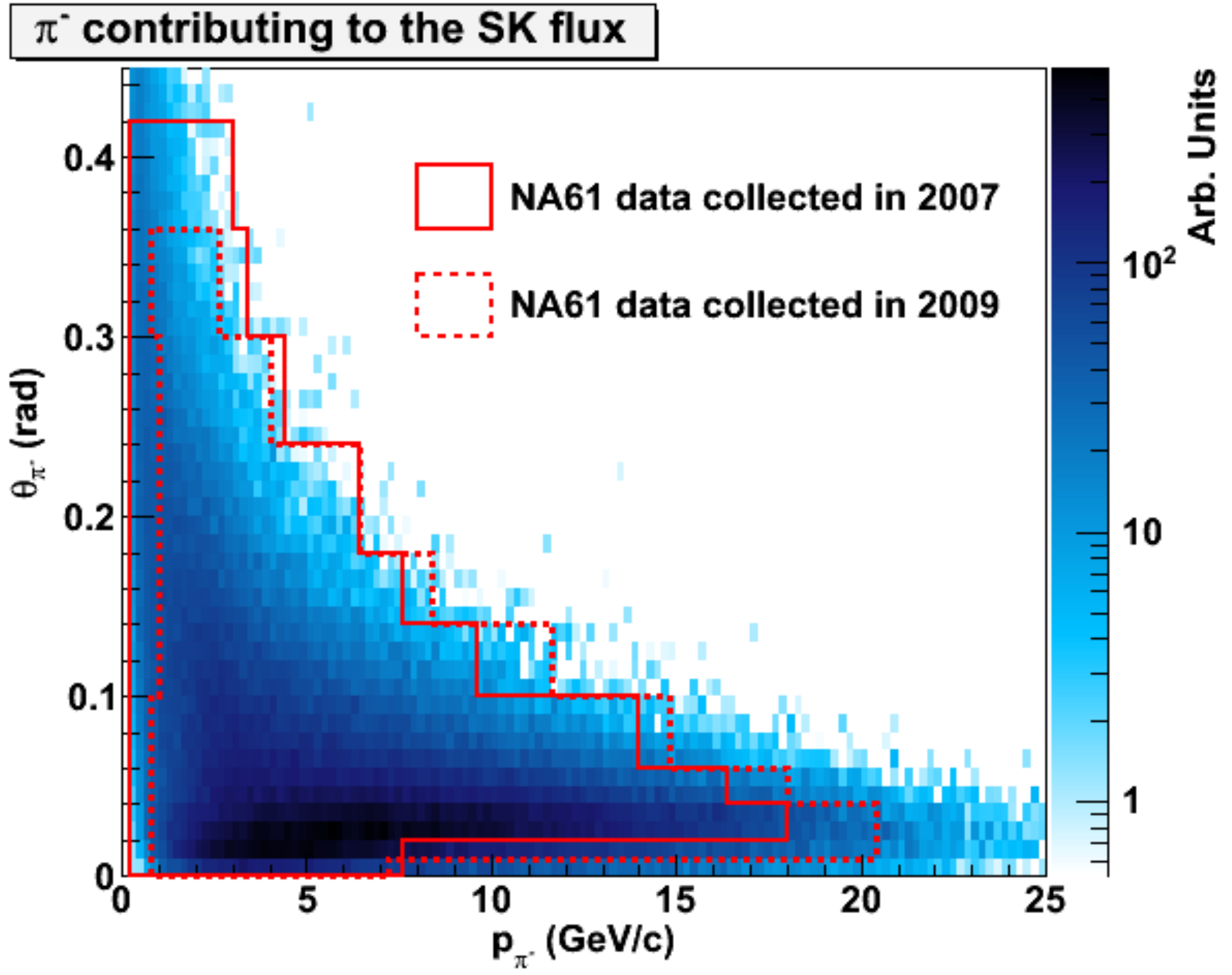}
\hspace*{0.4cm}
\includegraphics[width=0.26\textwidth]{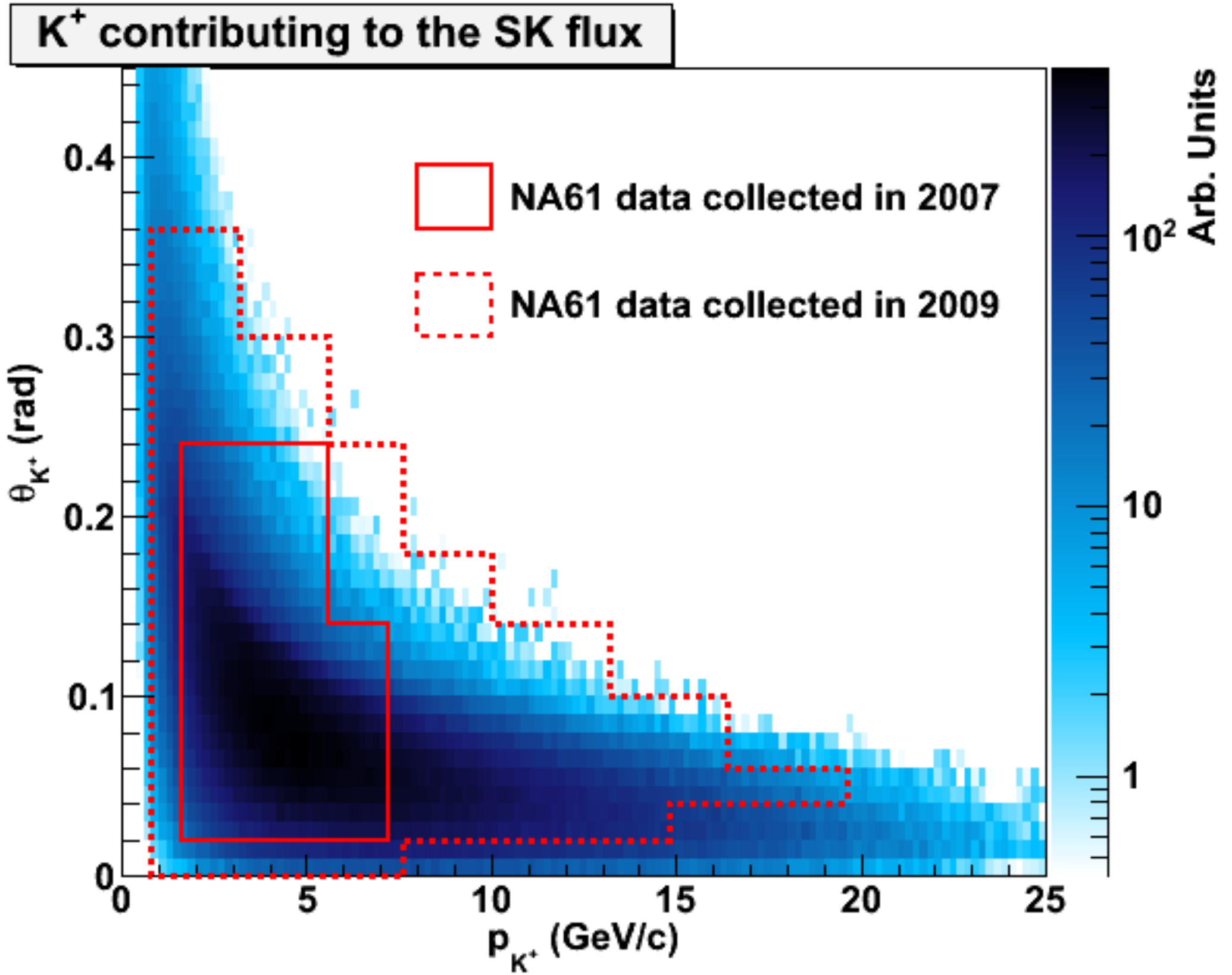}
\includegraphics[width=0.26\textwidth]{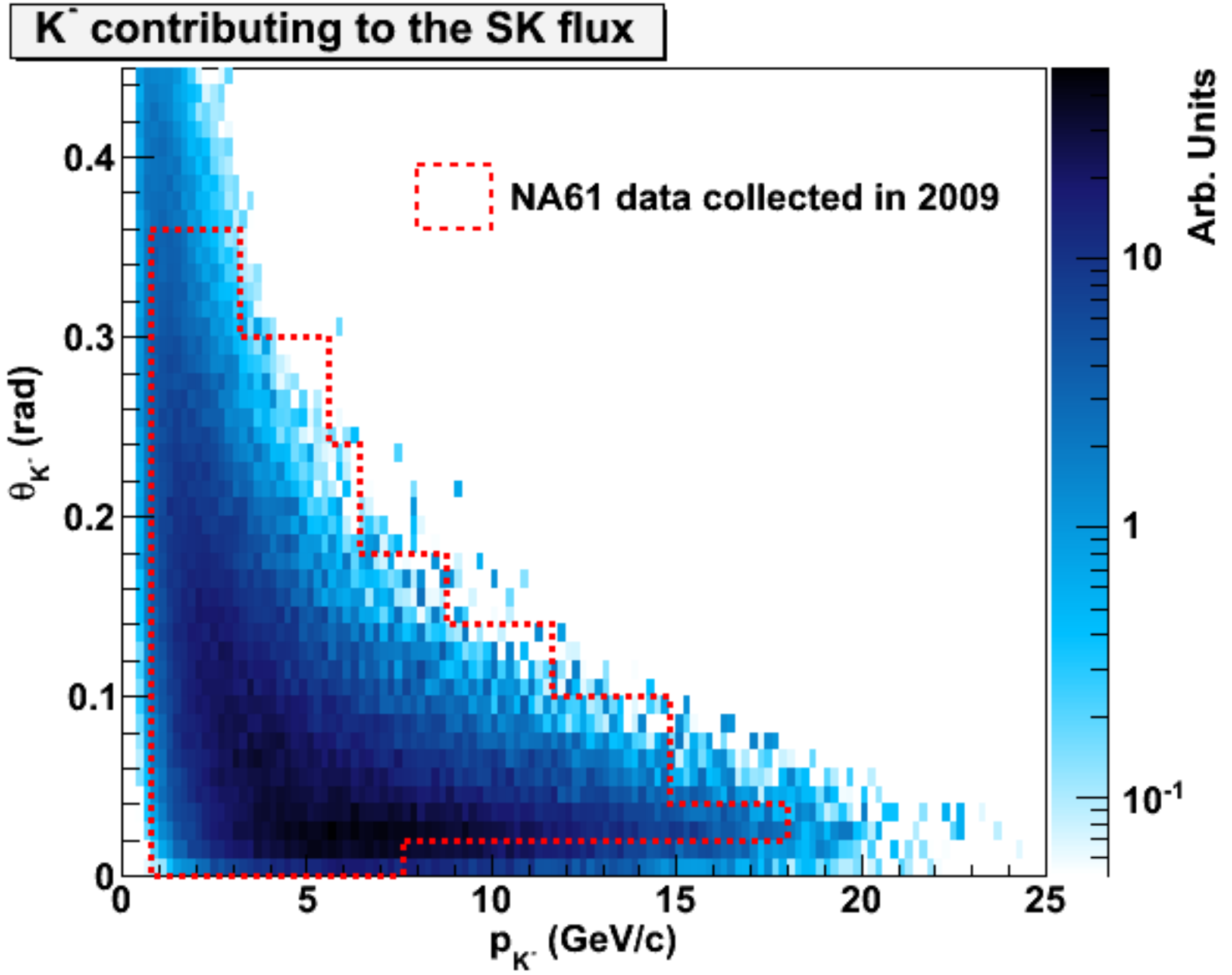}
\hspace*{0.4cm}
\includegraphics[width=0.26\textwidth,height=0.21\textwidth]{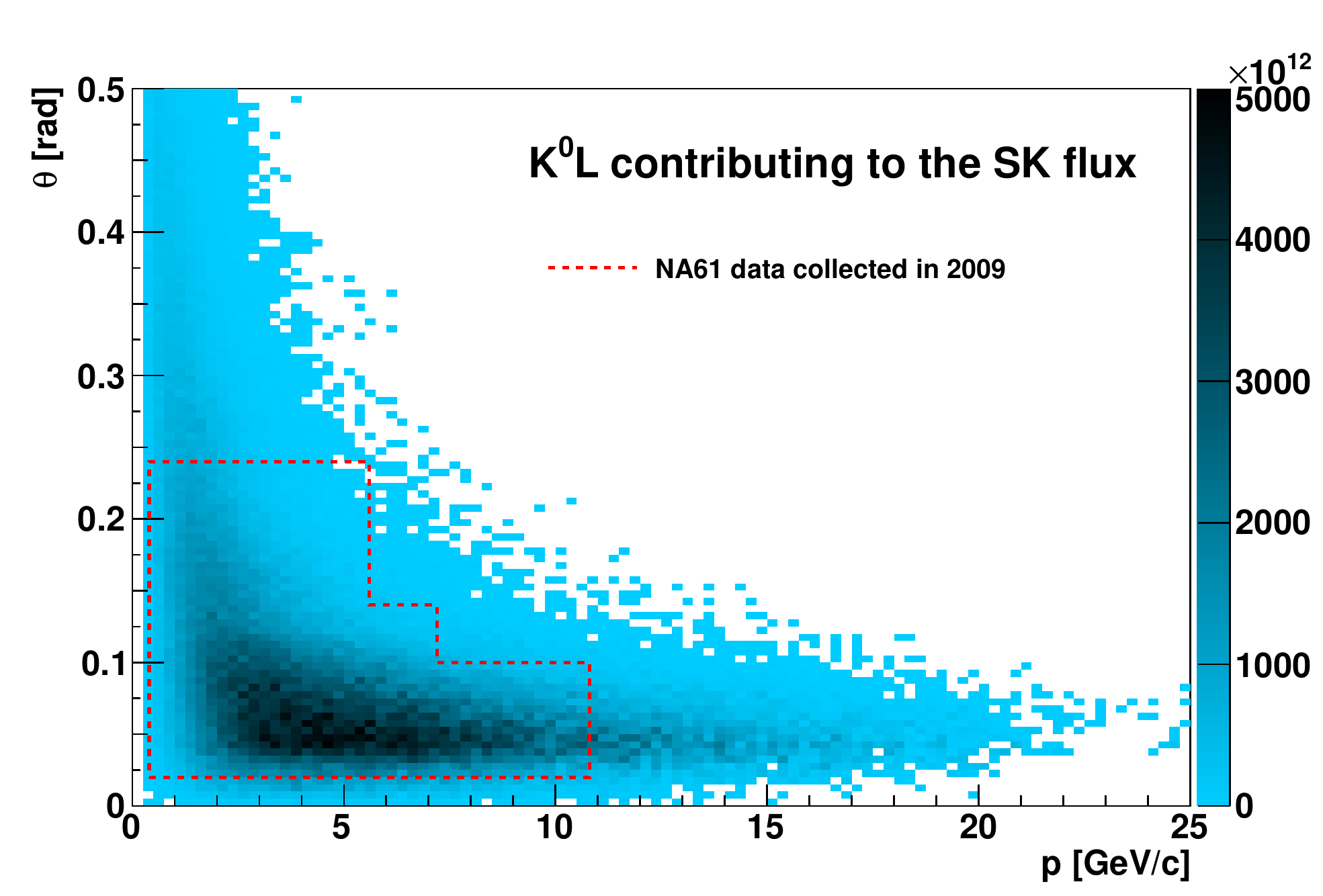}
\hspace*{0.4cm}
\includegraphics[width=0.26\textwidth]{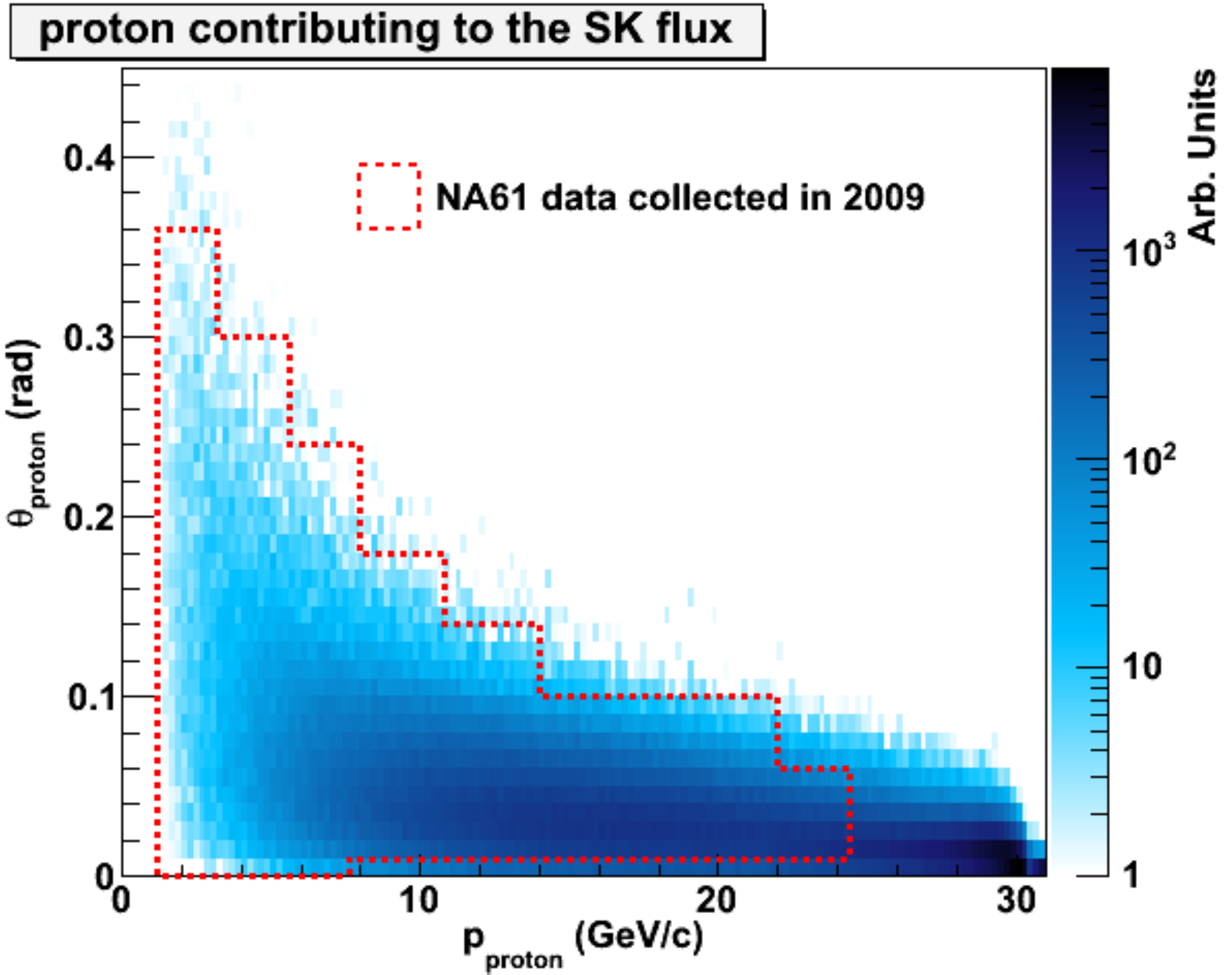}
\caption{The phase space of $\pi^+$, $\pi^-$, K$^+$, K$^-$, K$^0_L$
  and protons contributing to the predicted neutrino flux at SK 
  in the ``positive'' focusing configuration \cite{T2K_flux_paper}, 
  and  regions covered by the new 2009 data ({\it dashed line}) and by
   published NA61 data collected in 2007
  ({\it solid line}) \cite{Abgrall:2011ae,Abgrall:2011ts}.}
\label{fig:na61_coverage}
\end{figure}

Although pilot data 2007 covered a significant part of the relevant hadron production 
phase space of T2K \cite{T2K_flux_paper} the statistical uncertainty is quite large.
In the year 2008 important changes have been introduced to the experimental 
setup of NA61: new trigger logic, TPC read-out and DAQ upgrade,
additional sections of ToF wall, new beam-telescope detectors.
As a consequence of these upgrades the number of events recorded in 2009 and 2010
for about a same period of time have been increased by an order of magnitude
as compared to the 2007 run. This larger sample allows simultaneous extraction 
of yields of $\pi^{\pm}$, K$^{\pm}$, K$^0_s$ and protons.
Furthermore, the phase space 
of NA61 has been increased (Fig.\,\ref{fig:na61_coverage}).

\vspace{0.3cm}

{\it Normalization and production cross section.}
For the normalization of hadron spectra and the calculation of the production 
cross section we use a procedure described in \cite{Abgrall:2011ae}.
The idea is to
measure an interaction probability for cases when the graphite target
was inserted and removed. Using these values one calculates the so-called
``trigger'' cross section which, in turn, is an input for the analysis of
``physics'' cross sections.

\pagebreak

In the 2009 campaign the beam trigger ran simultaneously with the physics triggers.
The ability to apply the event-by-event selection improved the systematics significantly.
As a result of the analysis
\begin{wrapfigure}{r}{0.41\textwidth}
\centering
\vspace*{-0.1cm}
\includegraphics[width=0.41\textwidth]{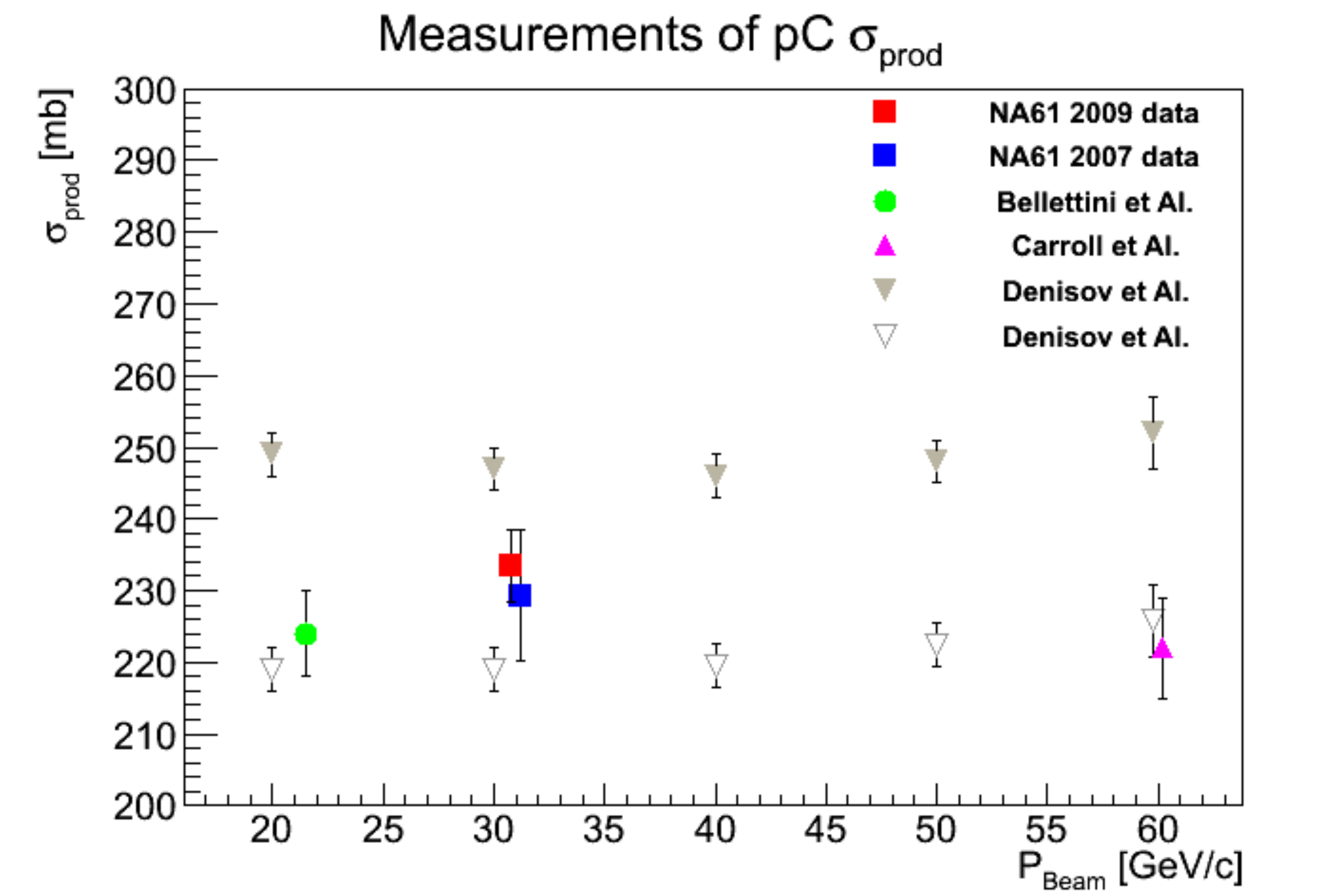}
\caption{A comparison of the production cross section to
  the previously published results obtained at different beam momenta
  \cite{Bellettini}.
}
\label{Davide}

\end{wrapfigure}
  of the 2009 data we obtain the production cross section
which comprises all processes due to strong \mbox{p + C} 
interactions excluding the coherent nuclear elastic and quasielastic scattering
\begin{eqnarray}
\sigma_{prod} = 233.5 \pm 2.8(\mbox{stat}) \pm 4.2(\mbox{model}) \pm 1.0 (\mbox{trig})~{\rm mb}.\nonumber
\end{eqnarray}
It is used further to normalize hadron cross sections to be able to compare to 
MC models.
All model-dependent corrections were estimated with GEANT4.9.5~\cite{GEANT4,GEANT4bis}
using FTF\_BIC physics list.
A comparison to the previously published results is presented in Fig.\,\ref{Davide}.

The total uncertainty of $\sigma_{prod}$ is 5.1 mb, almost a factor of two
smaller than the one obtained with the pilot 2007 data. 
The dominant contribution to the uncertainty comes from the physics model 
used to recalculate the production cross section from the ``trigger'' cross section.

\vspace{0.3cm}

{\it Measurement of charged hadron cross sections for T2K.}
Depending on the momentum interval and the particle type, different
analysis techniques were tested for the analysis of the 2007 data \cite{Abgrall:2011ae}. 
However because of larger statistics in 2009 it was decided to use only 
the combined analysis of time-of-flight (ToF) measurements and energy loss, $dE/dx$,
measurements in TPC \cite{Abgrall:2011ae} since this method provides a high selection 
purity of the sample and low dependence on a MC model.
The method can be applied for the region of momenta $p>1$ GeV/$c$, 
which is the most crucial for the T2K neutrino kinematics.

\begin{figure*}[]


\centering
\includegraphics[width=0.45\textwidth]{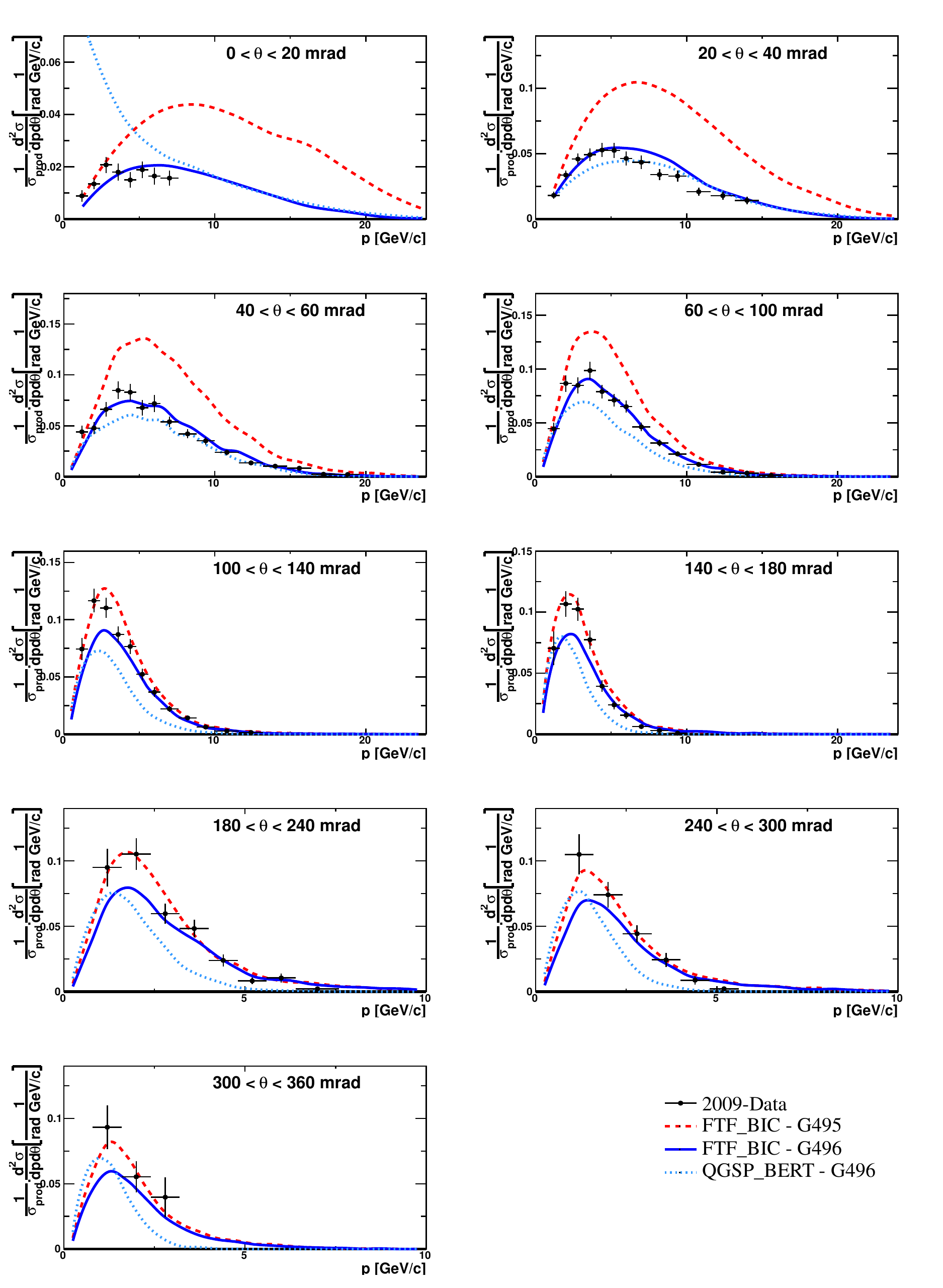}
\hspace{1cm}
\includegraphics[width=0.45\textwidth]{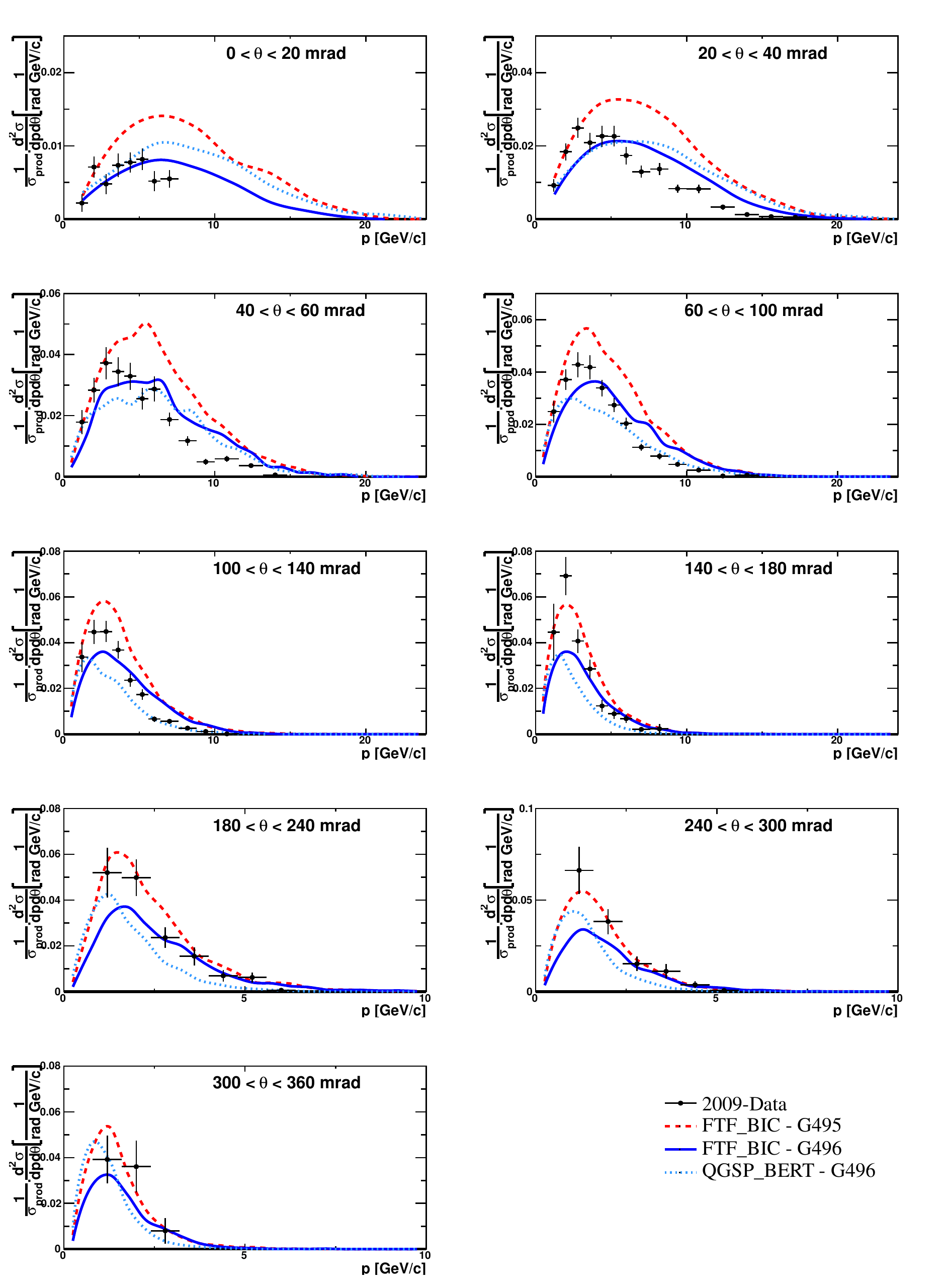}
\caption{Laboratory momentum distributions of the K$^+$ and K$^-$ multiplicities 
produced in p-C interactions at 31 GeV/$c$ in different intervals of polar angle $\theta$.
Error bars indicate statistical uncertainty.
Data points are overlapped by various model predictions \cite{GEANT4,GEANT4bis}. 
}
\label{NA61_kaons}
\end{figure*}

Comparison of pion spectra obtained with data 2007 and 2009 shows a good agreement 
\cite{SPSC_rep}.
In general the analysis of K$^\pm$ is more complicated 
due to a small fraction of kaons in the overall sample \cite{Abgrall:2011ts}. 
For instance, the K$^+$ signal vanishes over the predominant pion one 
at the low momentum range while at higher momenta protons dominate.
Statistical error of the K$^+$ spectra with the 2007 data 
was by a factor of 3 larger than the systematic one \cite{Abgrall:2011ts} 
and only two intervals in $\theta$ were considered.
Data collected in 2009 improved the precision strongly.
In Fig.\,\ref{NA61_kaons} the differential cross section of kaons 
normalized to the production cross section is shown. 
Larger statistics of 2009 allowed to split the phase space into 
9 intervals in $\theta$.
Several recommended GEANT4 physics lists 
are also presented on the figure \cite{GEANT4,GEANT4bis}.

\begin{figure*}[]
\begin{minipage}[b]{0.49\linewidth}
\centering
\includegraphics[width=0.9\textwidth]{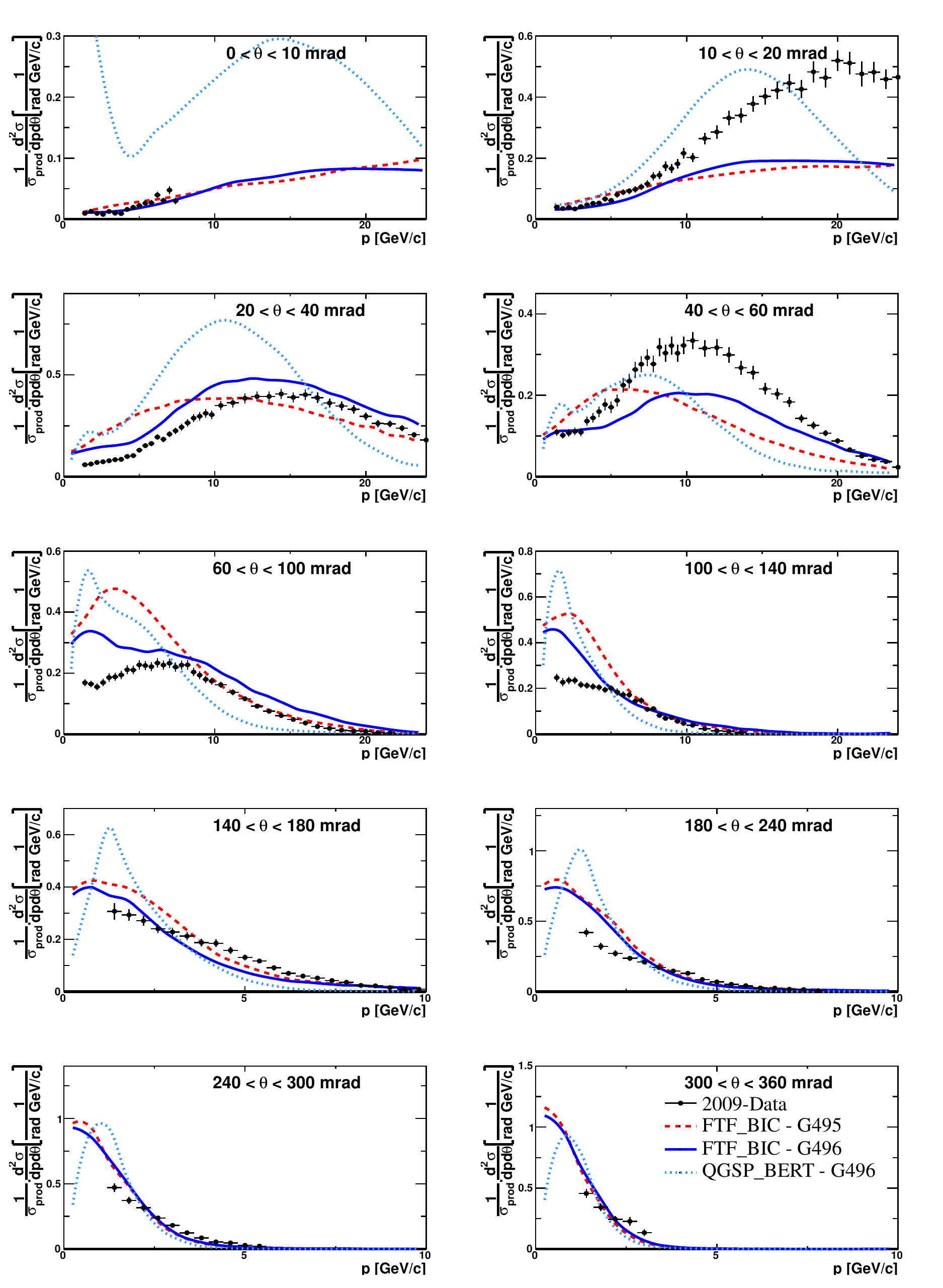}
\caption{
The same as Fig.\,\ref{NA61_kaons} but for protons.
}\label{NA61_protons}
\end{minipage}
\hfill
\begin{minipage}[b]{0.49\linewidth}
\centering
\includegraphics[width=0.9\textwidth]{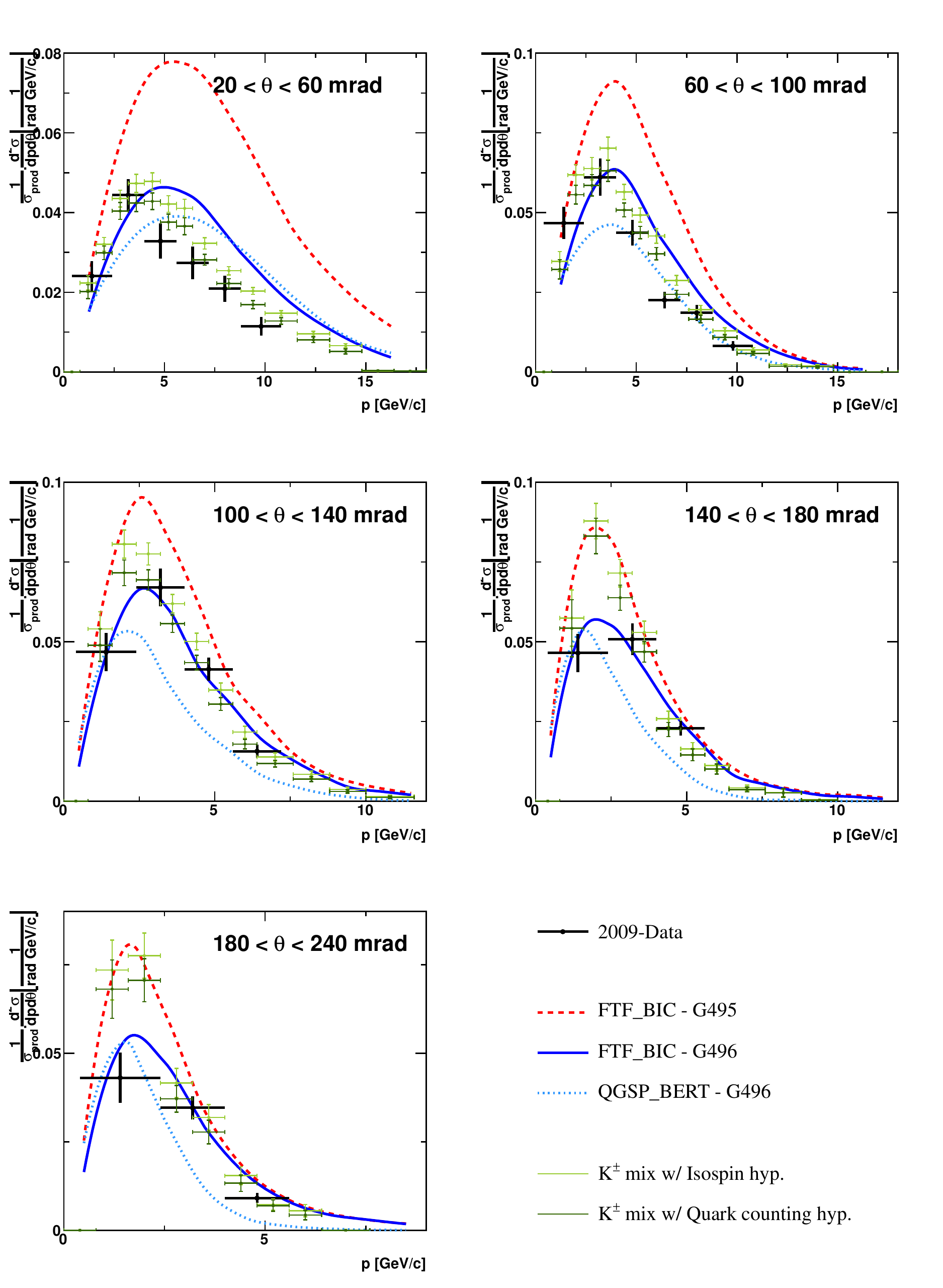}
\caption{
The same as Fig.\,\ref{NA61_kaons} but for $K^0_S$.
}
\label{NA61_K0S}
\end{minipage}
\end{figure*}

Distribution of proton multiplicities obtained with the 2009 data
as a function of momentum in different intervals of $\theta$ 
is shown in Fig.\,\ref{NA61_protons}.
Comparison to GEANT4 models shows that none of them 
can satisfactory describe the data.

\vspace{0.3cm}

{\it Cross section of neutral particles $K_S^0$ and $\Lambda$.}
Understanding of the neutral strange particle production
in NA61 allows to decrease the systematic uncertainty associated with 
the charged particle production, namely pions \cite{Abgrall:2011ae} and protons. 
In addition, the measurement of the $K^0_S$ cross section will improve 
our knowledge of the $\nu_e$ flux at the T2K experiment coming from 
the three body decay of the $K^0_L \to \pi^0 e^\pm \nu_e (\overline \nu_e)$ 
\cite{T2K_flux_paper}.
The technique used for the analysis of the 2009 data have been
already tested on the 2007 data \cite{K0_2007}.
$K_S^0$ and $\Lambda$ were reconstructed in the $V^0$ mode:
decay into two charged particles of opposite signs
$K^0_S \to \pi^+ + \pi^-$ and
$\Lambda \to p \phantom{^+} + \pi^-$.
Particle yields have been extracted in the analysis of invariant mass spectra
applying corresponding mass hypotheses for daughter tracks.
The main background sources were associated with
converted photons and the combinatorial background which is mainly 
due to particles produced in the primary interaction.
%
Multiplicities of $K_S^0$ are shown in Fig.\,\ref{NA61_K0S} together with 
a prediction of several models.
Analysis of $\Lambda$ is in progress.

Since yields of K$^+$, K$^-$ and K$^0_S$ could be measured simultaneously
they can provide a test for several hypotheses predicting a relative yield 
of charged and neutral kaons. 
%
%
%
The comparison of the measured differential multiplicity of K$^0_S$ 
to the prediction obtained with the charged kaons is shown in Fig.\,\ref{NA61_K0S}.
A reasonable agreement is observed.


\pagebreak

{\it Analysis of the T2K replica-target data.}
An essential fraction of the neutrino flux arises
from secondary re-interactions as long targets are used in T2K 
\cite{T2K_flux_paper}. 
The lack of direct data, hence the use of sparse data sets, to cover these 
contributions limits the achievable precision on the flux prediction. 
Therefore measurements with a full-size replica of the T2K target
have been performed by NA61 \cite{Abgrall:2012pp}. 

A dedicated reconstruction method has been developed to provide results 
in a form that is of direct interest for T2K.  Yields of
positively charged pions are reconstructed at the surface of the T2K
replica target in bins of the laboratory momentum and polar angle
as a function of the longitudinal position along the target. 
 By parametrizing hadron yields on a surface of the target
one predicts up to 90\% of the flux for both $\nu_{\mu}$ 
and $\nu_e$ components while only 60\% of neutrinos are coming from 
particles produced at the primary interaction vertex.
Two methods (constraint of 
hadroproduction data at primary interaction and on a target surface) 
are consistent within their uncertainties achieved on statistics of pilot run 2007 \cite{Abgrall:2012pp}.
The ultimate precision will come from the analysis of the replica-target data 
recorded in 2009 and 2010.

\end{document}